\documentclass[
preprint,
endfloats,
superscriptaddress,
nofootinbib,
amssymb,
aps,
pre
]{revtex4}

\usepackage{amsmath}
\usepackage{graphicx}

\newcommand{\xpl}{x_{\text {pl}}}

\newcommand{\Vsat}{V_{\text {sat}}}
\newcommand{\eff}{\epsilon_{\text {eff}}}

\begin{document}

\title[Cochard]{Stabilization of frictional sliding by normal load modulation:\\
A bifurcation analysis}

\author{A. Cochard} 
\affiliation{Laboratoire de D\'etection G\'eophysique, CEA\\
B.P. 12, 91680 Bruy\`eres-le-Ch\^atel, France}
\affiliation{Laboratoire de G\'eologie (UMR 8538),\\
\'Ecole Normale Sup\'erieure,\\
24, rue Lhomond, 75231 Paris, Cedex 05, France}
\author{L. Bureau}
\affiliation{Groupe de Physique des Solides (UMR 7588),\\
Universit\'es Paris 6 \& 7,\\
2, place Jussieu, 75251 Paris, Cedex 05, France}
\author{T. Baumberger}
\email[To whom correspondance should be 
addressed --- email: ]{tristan@gps.jussieu.fr}
\affiliation{Groupe de Physique des Solides (UMR 7588),\\
Universit\'es Paris 6 \& 7,\\
2, place Jussieu, 75251 Paris, Cedex 05, France}

\date{\today}

\begin{abstract}
This paper presents the stability analysis of a system sliding at low 
velocities  ($< 100\,\mu$m.s$^{-1}$) under a
periodically modulated  normal load, preserving interfacial contact. 
Experiments clearly evidence that 
normal vibrations generally stabilize the system against stick-slip 
oscillations, at least for a modulation frequency much larger than 
the stick-slip one. 
The mechanical model of Bureau {\it et al.} (2000), validated on the
steady-state response of the system, is used to map its stability diagram.  
The model takes explicitly into account the finite shear stiffness
of the load-bearing asperities, in addition to a classical
state- and rate-dependent friction force.
The numerical results are in excellent quantitative agreement with the
experimental data obtained from a multicontact frictional system
between glassy polymer materials. 
Simulations at larger  amplitude of modulation (typically $20\%$ of 
the mean normal load) suggest that the non-linear coupling 
between normal and sliding motion could have a destabilizing effect 
in restricted regions of the parameter space.  
\end{abstract}
\maketitle
\section{Introduction}

It is well known that non-linearities in the constitutive
laws of dry friction may lead to the instability of steady frictional
sliding against stick-slip oscillations, even for a single degree-of-freedom
system driven at nominal constant velocity through a compliant
stage.
Sliding instability is an important issue in mechanical engineering since
it is an ultimate limitation to the positioning accuracy for
precision structures and machine tools. 
Moreover, stick-slip oscillations may be
strongly non-linear and make servo-control extremely difficult.
When designing a sliding
mechanism, it is therefore of primary importance
to characterize accurately the variations
of the friction coefficient with, {\it e.g.}, sliding velocity, keeping
in mind that even slight variations may have a destabilizing effect.
This requires to go beyond Amontons-Coulomb's law which assumes a constant
friction coefficient. It might be legitimately feared that a more detailed
constitutive law would have a restricted scope, {\it e.g.}, in
terms of materials and range of sliding velocities.
It is thus remarkable that in the limit of low
velocities (typically lower that $100\,\mu$m.s$^{-1}$),
and light enough
loads so that the interface is made of a sparse set of micro-contacts
between load-bearing asperities, a relatively material-independent
frictional behaviour is found which can be accounted 
for by a simple set of non-linear constitutive equations.
Such studies have been initiated in the
field of rock mechanics by
Dieterich \cite{DIET} and
Rice and Ruina \cite{RR}, motivated by the need for low velocity
friction models
to investigate fault dynamics and earthquake nucleation.
They have put on a firm phenomenological basis
the idea, already suggested by the work of Rabinowicz \cite{RAB}, that
friction does not depend only on the instantaneous sliding velocity
$v$
but also on the whole sliding history.
An experimental signature is the hysteretic frictional response of the
interface when the slider is driven at a non-steady rate.
Rice and Ruina \cite{RR} proposed a
family of dynamical equations coupling the sliding velocity to a set
of state variables.  Subsequent experimental investigations
have shown that a single state variable $\phi$ is sufficient for most
purposes. These experimental studies were performed on a wide range of
materials,
such as granite \cite{DIET},
paper \cite{PRE94}, polymer
glasses \cite{PRB2}
and elastomers \cite{PRS01}.
The friction force in this model is $F = W\,\mu(v, \phi)$
with $W$ the normal load and $\mu$ the friction coefficient.
Moreover, it has been
possible \cite{DK,PRB1} to give a physical
interpretation of $\phi$ as the average ``age'' of the
microcontacts 
which grow while the material creeps
under normal load, until sliding interrupts the process by renewing
the load-bearing contact population.
The dynamical model is closed by specifying a differential equation
coupling $v$ to $\phi$ so as to account for the renewal of
the microcontact population after a slip length $D_{0}$ of micrometric
order. This length is of order the mean radius of the microcontacts between
surfaces of micrometric roughness \cite{GW}.
The resulting state- and rate-dependent
friction laws will be hereafter 
referred to as SRF.  
Among several SRF expressions proposed originally, the one that we use in this 
paper are:

\begin{equation}
\label{eq:frico}
\mu(v, \phi)=
\mu_0 
+ A\ln\left(\frac{v}{V_0}\right) 
+ B\ln\left(1+\frac{\Vsat\,\phi}{D_0}\right)
\end{equation}
for the friction coefficient and
\begin{equation}
\label{eq:phidot}
\frac{d\phi}{dt} =1-\frac{v\phi}{D_0}
\end{equation}
for the evolution of the state variable, where $\mu_0$, $A$, $B$, 
$Vsat$ and $V_0$ are constants.

This SFR model has been extensively validated by testing 
against numerous experimental situations involving  transient dynamical 
responses of the system. 
The most stringent test
relies upon the non-linear characteristics of the bifurcation from
steady-sliding to stick-slip oscillations \cite{PRE95}.
The model can be understood 
as resulting from two distinct physical mechanisms, the effect of 
which can be summarized in the following decomposition of the friction 
force, proposed by Bowden and Tabor \cite{BT}, in terms of the real area of contact
$\Sigma_{r}$,
and an interfacial shear strength $\sigma_{s}$:

\begin {equation}
\label{eq:taborsrf}
F(v, \phi)  = \sigma_{s}(v) \Sigma_{r}(\phi)\ .
\end{equation}

Here, the real area of contact depends on the interfacial age  because 
it grows due to the creep of the load-bearing asperities \cite{PRB1}. The velocity 
dependent interfacial shear strength has been ascribed to the 
adhesive, nanometer thick  junctions between microasperities. A simple
microscopic model has been proposed for 
the elasto-visco-plastic rheology of the junctions,
compatible with the existence of a finite friction threshold 
\cite{PRB2}. 
 
Recently, attention has been paid to the effect of a time-dependent normal
load on the response of a single degree-of-freedom sliding system.
This situation is of practical interest when the mechanical design
allows cross-talking between the normal
(loading) and the tangential (driving) forces \cite{DB},
or when external vibrations
contribute to the loading of the interface, as it may be the case for 
seismic
faults \cite{LD,PERF,RM}.
The response of the system is not intuitive.
First, since the friction force is directly proportional to the
normal load, the sliding velocity is dynamically coupled to the normal load
modulation, hence feeds back the friction force.
Moreover, it has been shown that more subtle interplays must be taken
into account.
Linker and Dieterich \cite{LD} have interpreted the transient
response to a step in normal load by coupling directly the time-variations
of $\phi$ and $W$, thus adding a term
$-\text{const}\,\phi\,d\ln(W)/dt$ 
in Eq. (\ref{eq:phidot}). The physical
motivation for this extension of the 
SRF ageing equation is the fact that, according to \cite{GW}, a change in normal force creates fresh load-bearing
contact area. This certainly influences directly
the age $\phi$, though probably in a weaker measure than
proposed by Linker and Dieterich \cite{LD}, as briefly discussed in
\cite{PRE00}.
More recently, Bureau et al. \cite{PRE00} have studied the response of
a sliding system to a periodically modulated normal load
$W(t) = W_{0}[1 + \epsilon \cos(\omega t)]$ with $\epsilon < 1$.
They found that
the friction force, averaged on a modulation period, is significantly lowered
with respect to the
situation under constant load $W_{0}$. The oscillating part of the force,
primarily reduced to its spectral component at $\omega$ in the limit of
vanishing $\epsilon$,
becomes quickly anharmonic as $\epsilon$ is increased while still remaining
much smaller than 1. They have shown that the SRF equations can
fit accurately
all their results provided that the model is modified to  account for
the finite interfacial shear stiffness $\kappa$ 
resulting from the elastic deformation of the
load-bearing asperities. 
This means that
the sliding
velocity differs from the velocity of the center of mass of the slider,
a statement which is
clearly illustrated in the static state, i.e., for tangential forces
well below the static
threshold, where the interface responds elastically without
sliding \cite{PRS,RSI}.
Under constant normal load and constant driving velocity $V$, this ``hidden'' interfacial degree of freedom
manifests itself only
for non-steady motion, and plays no significant
role at the circular frequency $\Omega_{c} \sim V/D_{0}$ of the
oscillations at the onset of the stick-slip instability.

However, under a modulated load at $\omega \gg
\Omega_{c}$, one must take the finite interfacial compliance $\kappa$
into account, all the more so since the latter is known to
be itself
proportional to $W$ \cite{PRS}.  This
results in a non-trivial and efficient coupling between the normal load
and the sliding velocity.

Of particular interest is the effect of load modulation on the
sliding stability of the system.
Dupont and Bapna \cite{DB} have computed the critical stiffness of the drive
below which stick-slip occurs for a slider-spring loaded at a
constant angle with respect to the sliding plane. This configuration
would provide a direct test for the coupling between $\phi$ and $W$
proposed by Linker and Dieterich \cite{LD}, but the experimental study has
not been performed so far.

The present paper addresses the problem of the stability of a slider-spring
system under an externally and harmonically modulated normal load. The
experimental arrangement is described in section 2 and it is shown
that for a circular frequency
$\omega \gg \Omega_{c}$ the modulation generally
stabilizes the system against stick-slip.
This spectacular effect is accounted for by the SRF model with
modulated interfacial stiffness, as shown by the numerical study of the
bifurcation which is detailed in section 3.

\section{Experiments}

\subsection{Apparatus}

\begin{figure}
\includegraphics{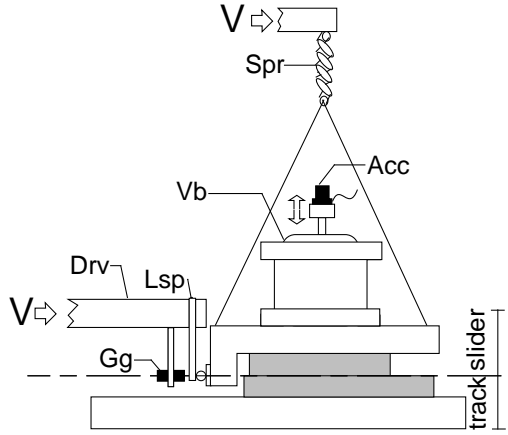}
\caption{Main elements of the experimental setup: 
Translation stage (Drv); Loading leaf spring (Lsp); Displacement 
gauge (Gg); Vibration exciter (Vb); Weighting spring (Spr);  
Accelerometer (Acc).}
\label{fig:setup}
\end{figure}

The apparatus (Fig.~\ref{fig:setup}) consists of a slider of mass $M$
driven along a track through a loading spring of stiffness $K = 0.21$
N.$\mu$m$^{-1}$. The
loading end of the spring is moved at constant velocity $V$ in the
range 1--100 $\mu$m.s$^{-1}$ by means of a translation stage driven by
a stepping motor.
The spring elongation is measured by an inductive probe
(Electro, sensor 4937, module
PBA200), with a 0.1 $\mu$m resolution over the 10 kHz bandwidth.
The average normal load $W_{0}$ can be set in
the range 3--23 N with the help of a vertical spring attached
to a remote point itself translated horizontally at the pulling
velocity $V$ through a second translation stage, in order to
prevent any tangential coupling.
The normal load modulation is achieved by means of a
vibration exciter (LDS, model V100) rigidly attached to the slider: a harmonic
voltage input of given amplitude and frequency $f$ results in a
harmonic vertical motion of the moving element of the exciter
on which an accelerometer (Br\" uel \& Kj\ae r, type
4371 V) is fixed. An acceleration of
amplitude $\gamma$ of this moving element of mass $m$ induces a
normal load modulation on the slider of amplitude $m\gamma$ at
frequency $f$. We thus obtain a normal load
$W(t) = W_{0}[1 + \epsilon \cos(\omega t)]$
with $\omega = 2\pi f$ and $\epsilon = m\gamma/W_{0}$ in the range
0.01--0.5. A fixed frequency $f = 120$ Hz has been used
for the whole study.
Two poly(methylmethacrylate) (PMMA) samples are glued,
respectively, on the slider and the track.  They have nominally flat surfaces
which have been
lapped together with 400-grit SiC powder and water to obtain
a rms roughness $R_{q}$ = 1.3 $\mu$m \cite{PRS}.
The interface between the two blocks is made of a sparse set of
load-bearing microcontacts \cite{PRS}. An air layer of
micrometric thickness is therefore trapped between the surfaces
and acts as a viscoelastic element, in
parallel with the microcontacts,
which  partially bears the normal load. This effect has been studied
in details in Bureau et al. \cite{PRE00} who concluded that the remaining effect
of the load
modulation on the asperities can be described by an effective
amplitude $\eff = \rho \epsilon$, with $\rho$ a constant close to 0.5, 
taken in the 
following as $\rho = 0.43$, a value which will be 
justified in section~\ref{sec:gigafit}.

\subsection{Localization of the stick-slip bifurcation, effect of the 
modulation}

The bifurcation between steady sliding and stick-slip oscillations
under constant load ($\epsilon = 0$) has been extensively described (see
{\it e.g.} \cite{PRE94}). When $K$ and $V$ are kept constant, 
steady sliding occurs for values of the remaining control
parameter $W_{0} < W_{0}^c(V)$ where the $K$-dependency has been 
omitted here since the value of $K$ is fixed in this study. 
The bifurcation is of the direct Hopf kind,
which means that
the amplitude of oscillation of the slider velocity is
vanishing
\footnote{Note that the term ``slick-slip''
is therefore a misnomer
since the
sliding velocity does not reach zero, {\it i.e.}, the slider does not
``stick'' during an oscillation period.}
when
approaching $W_{0}^c$ from below ({\it i.e.} increasing $W_{0}$ up to $W_{0}^c$),
while the circular
frequency tends to a finite value $\Omega_{c}$.
In addition, the characteristic time of the oscillating transients diverges
when approaching $W_{0}^c$ from above ({\it i.e.} decreasing $W_{0}$ 
towards $W_{0}^c$).
A practical consequence is that
as the bifurcation is tracked down,
it becomes increasingly difficult to distinguish between steady stick-slip
oscillations and transient relaxation towards steady sliding, 
resulting from
the perturbating effect of friction force fluctuations along the track.

\begin{figure}
\includegraphics[width = 10 cm]{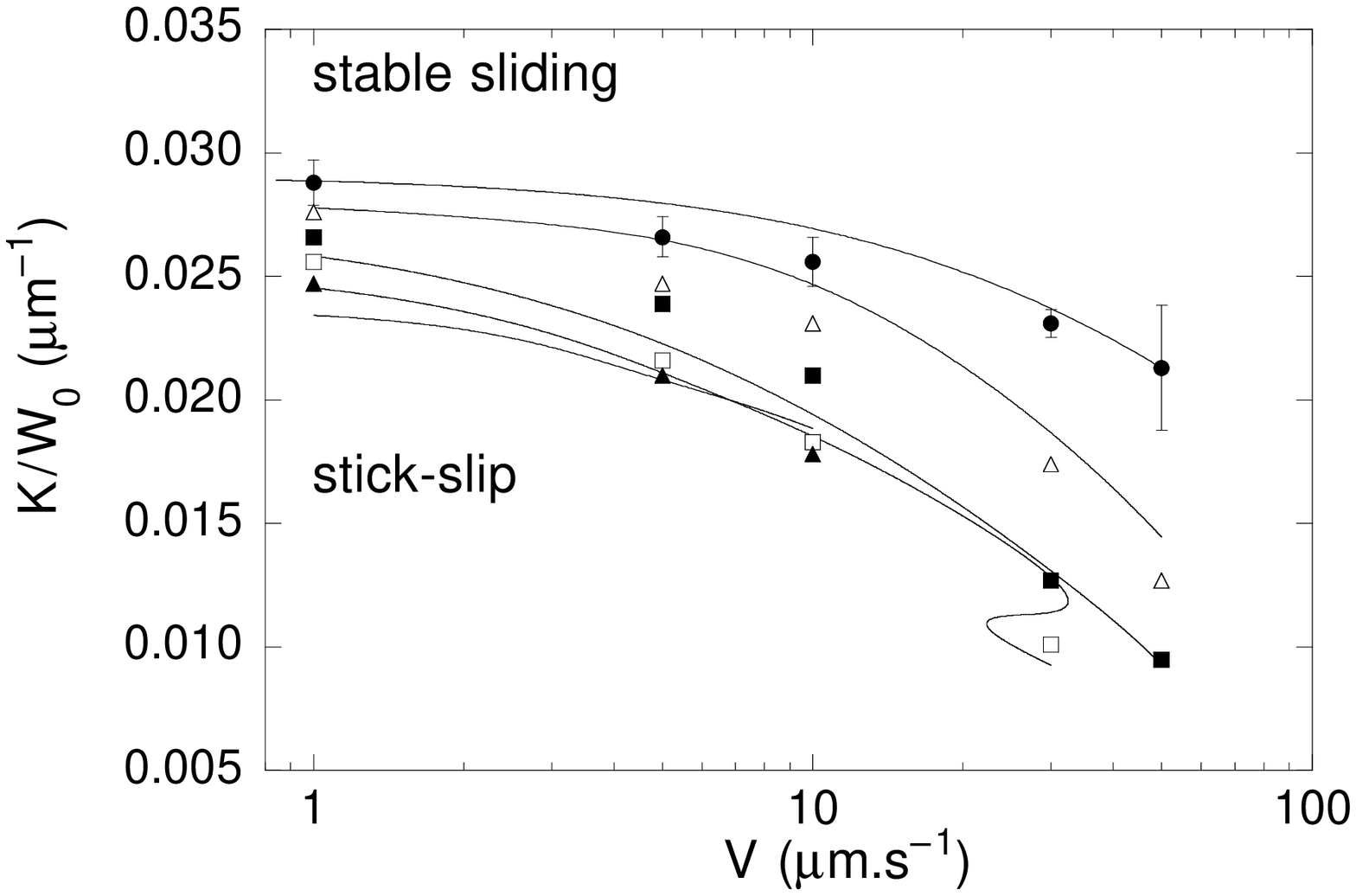}
\caption{
Stability diagram for different values of the modulation 
amplitude. For given $V$ and $\eff$, bifurcation from stick-slip to stable 
sliding occurs when the control 
parameter $K/W_{0}$ overcomes the plotted critical value: 
$\eff = 0\, (\bullet)$; 
$0.045\, (\triangle)$;
$0.09\, (\blacksquare)$;
$0.13\, (\square)$;
$0.18\, (\blacktriangle)$.
For the sake of clarity, typical standard deviations are plotted as error bars  only 
for  $\eff = 0$.
The solid line curves are the output of the numerical study (see 
section~\ref{sec:gigafit}). The larger $\eff$ the lower the curve at $V = 1\mu$m.s$^{-1}$. 
}
\label{fig:stability}
\end{figure}

For $\epsilon = 0$, the ratio
$K/W_{0}$ is the relevant control parameter, at least in the low velocity
region where
inertia of the slider oscillating at the 
circular frequency $\Omega_{c}$ can be neglected  \cite{PRE94}. 
Henceforth, although the external
stiffness $K$ is kept
constant for the whole set of experiments reported in this paper,
we will 
keep on representing the stability domain of the system
in the parameter plane $(K/W_{0}, V)$ where it is bounded by the experimental
bifurcation curve (Fig.~\ref{fig:stability}). 
The  experimental uncertainty on the critical value $K/W_{0}^c$,  
determined from the standard deviation over at least 10 measurements
at a given velocity $V$, is indicated by the error bars in Fig.~\ref{fig:stability}.
It  is typically $\pm 3\%$, except at the larger velocity, since the
results are more sensitive to long 
wavelength irregularities along the track for large sliding distances.

When a harmonic modulation at $\omega$ is superimposed to a value of $W_{0}$
corresponding to steady sliding at $V$, the velocity of the slider oscillates
about $V$, possibly in a anharmonic way, with
a fundamental component at $\omega$. The motion of the slider, when
averaged over a period $2\pi/\omega$ is therefore steady.
For given $V$ and $K$, one has now to consider two control 
parameters, namely $W_{0}$ and $\eff$.

\begin{figure}
\includegraphics[width=10cm]{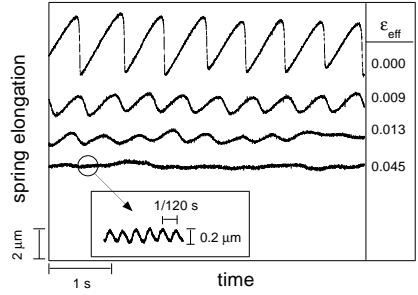}
\caption{
Time evolution of the loading spring elongation for $V = 8\,
\mu$m.s$^{-1}$ and different modulation amplitudes $\eff$ indicated at 
the right end of each trace. A vertical offset 
has been added to each trace in order to display clearly the 
bifurcation sequence 
from stick-slip to stable sliding. The inset is a blow-up of the stable 
sliding trace showing the remaining oscillating response at the 
frequency of the load modulation ($f = 120\,$Hz, much higher than 
the stick-slip frequency).  
}
\label{fig:trace}
\end{figure}

When the d.c. load $W_{0}$ is increased while keeping constant
$\eff$, the {\it average} slider motion
is found to become of the stick-slip kind above a critical value, 
$W_{0}^{c}(V,\eff)>W_{0}^{c}(V)$ (Fig.~\ref{fig:stability}).
A normal load modulation of even very small effective amplitude
may therefore {\it stabilize} the system
against stick-slip as illustrated directly in Fig.~\ref{fig:trace} 
where a modulation with $\eff = 4.5\times 10^{-2}$ is enough to suppress 
well developed, strongly anharmonic, large amplitude and low frequency 
stick-slip oscillations (the force signal then only shows the 
the remaining small amplitude modulation at the forcing higher frequency). 

The effect of $\eff$ on the critical value of $K/W_{0}(V)$ is 
characterized in Fig.~\ref{fig:chichi}. The higher the velocity, the 
stronger the stabilizing effect of the normal load modulation. 
The effect is  spectacular when described in terms of the
velocity domain corresponding to steady sliding at constant
$K/W_{0}$. For instance, the critical velocity at $K/W_{0}=0.026
\,\mu$m$^{-1}$ is decreased by more than a factor of ten by applying a
modulation with $\eff=0.09$.

The empirical study indicates that, as a rule of thumb, 
steady-sliding is promoted by high velocity, high amplitude of normal load 
modulation, low average normal load and large stiffness. This is tested in the 
following against a numerical study of the SRF model including normal 
load modulation. 
 
\section{Numerical study}

\subsection{The state- and rate-dependent fiction (SRF) model equations}

\begin{figure}
\includegraphics{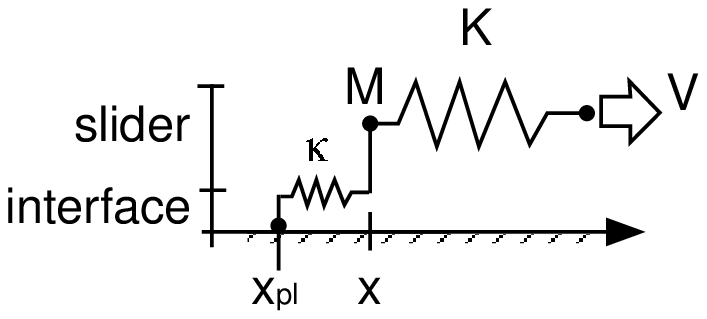}
\caption{
Equivalent mechanical circuit of the slider/track system. $K$ 
is the stiffness of the loading spring, $\kappa$ is the one of the 
interface. 
}
\label{fig:kinematic}
\end{figure}

The SRF laws (Eqs.~\ref{eq:frico}, \ref{eq:phidot}) are 
incorporated into the equation of motion of the slider, according to 
the simple model sketched in Fig.~\ref{fig:kinematic}. 
The proportionality constant between the normal load $W$ and the
interfacial stiffness $\kappa$ is a length $\lambda$ of 
micrometric order: 
$\kappa=W/\lambda$ \cite{PRS}.  
The equation of motion of the
slider thus reads:
\begin{equation}
\label{eq:newt}
M x'' = K (V t -x) - \kappa(x-\xpl) 
\end{equation}
where here and henceforth the prime denotes time derivative.  
Taking a massless interfacial zone (a reasonable
assumption, see appendix), we also have, according to Eq. 
(\ref{eq:frico}): 
\begin{equation}
\label{eq:newt0}
\frac{W}{\lambda} (x - \xpl) = W \left[\mu_0 
+ A\ln\left(\frac{v}{V_0}\right) 
+ B\ln\left(1+\frac{\Vsat\,\phi}{D_0}\right)\right] \ ,
\end{equation}
where $v=x'_{\text {pl}}$ is the relative sliding velocity at the
interface and
$\phi$ follows the evolution law (\ref{eq:phidot}) rewritten here for 
the sake of clarity
\footnote{This equation, including the actual sliding velocity $v$ in 
place of $x'$ is consistent with our physical understanding of the 
age $\phi$. It has been checked that mistaking $x'$ for $v$, 
as in \cite{PRE00},
has no 
significant effect on their results, at 
least for the modulation frequencies much larger than $V/D_{0}$ used 
in their study.}
:
\begin{equation}
\label{eq:phidot2}
\phi'(t)=1-\frac{v\,\phi(t)}{D_0} \ .
\end{equation}
 
For numerical purposes, we wish to recast those equations in the form
of a system of first order ordinary differential equations (ODEs).  
Noting $z=x'$,
$u=V\,t-x$, further differentiating $x-\xpl$ with
respect to time in (\ref{eq:newt0}) using the explicit expression for
$W(t)=W_0[1+\eff\cos(\omega\,t)]$ and solving for $v'$, we get the
following ODE system, which we will use for the numerical bifurcation
analysis:

\begin{widetext}
\begin{eqnarray}
\label{eq:dotu}  
u' & = & V-z \\
\label{eq:dotz}  
z' & = & \frac{K}{M} u\nonumber\\
 &&- \frac{W_0(1+\eff\cos(\omega\,t))}{M} \left[\mu_0 + A \ln\left(\frac{v}{V_0}\right) 
+ B \ln\left(1+\frac{\Vsat\,\phi}{D_0}\right)\right] \\
\label{eq:dotv} 
v' & = &\frac{v}{\lambda\,A}
\left[z-v-\frac{\lambda \, B \, \Vsat (1-v\,\phi/D_0)}{D_0(1+\Vsat\,\phi/D_0)}
\right] \\
\phi' & = & 1-\frac{v\,\phi}{D_0} \ .
\end{eqnarray}
\end{widetext}

\subsection{Determination of the SRF parameters}
\label{sec:deterpar}

In order to analyse the data within the SRF framework, we need to
determine a set of values for the relevant parameters of the model.
This is performed under constant normal load, according to a  well
established procedure. The values, which will be used in the numerical
analysis, some of them as trial ones, are gathered in Table~\ref{tab:par}.
The useful formulae are established in the appendix. 

\begin{enumerate}
\item[(i)] First, the steady sliding friction coefficient $\mu_{d}(V)$ is
measured to be velocity-weakening with an almost constant logarithmic slope
over the 1 -- 100 $\mu$m.s$^{-1}$ range. This indicates that
$\Vsat$, above which $\mu_d(V)$ increases with increasing $V$ 
according to Eqs.~(\ref{eq:frico}, \ref{eq:phidot}), 
is certainly larger than 100 $\mu$m.s$^{-1}$ and allows to extract
a value for $B-A$ and $\mu_{0}$ at $V_{0} = 1\,\mu$m.s$^{-1}$, {\it i.e.} 
far below the saturation of the ageing term.

\item[(ii)] Next, the critical value of $K/W_{0}$, here taken at midrange (10 
$\mu$m.s$^{-1}$) where inertial terms can be neglected in 
Eq.~(\ref{eq:kcinert}),
yields a determination of the memory length  $D_{0}$.

\item[(iii)] From the value of the critical stick-slip period at 10
$\mu$m.s$^{-1}$ (Eq.~(\ref{eq:pulsinert})), we obtain a value for $A$.

\item[(iv)] A determination of $\Vsat$ is finally obtained by a best fit of the whole
bifurcation curve in the plane $(K/W_{0}, V)$ for $V$ in the
1 -- 50 $\mu$m.s$^{-1}$ range, treating the inertial term in 
Eq.~(\ref{eq:kcinert}) as a perturbative one. Since the value of 
$\Vsat$ is out of the experimental velocity window, this 
determination is not very accurate. Treating several data set 
corresponding to different runs yields an uncertainty  as large as $\pm 25 \%$
on the value of $\Vsat$. 

\item[(iv)] The value of the length $\lambda$, defined by the ratio of the 
load $W_{0}$
and the interfacial shear stiffness $\kappa$, has been obtained in
\cite{PRE00} from a best fit of the a.c. response of the slider
position to the normal load modulation.

\end{enumerate}

It is clear that this procedure, though systematic, generates cumulative 
errors which are difficult to evaluate (the uncertainties on $A$, $B$ 
and $D_{0}$ given in
Table~(\ref{tab:par}) are conservative values). In view of the high sensitivity 
of the bifurcation  to small variations of the parameters,  
we have chosen in the following numerical analysis to use the set 
determined above as a trial one. Namely, the parameters which are 
left free are $A$, $B$, $D_{0}$, $\lambda$, $\Vsat$, and the 
ratio $\rho = \eff/\epsilon$.  

\subsection{Bifurcation analysis}
\label{sec:gigafit}

Technically, the transition from steady sliding to stick-slip, both
states being modulated by the forcing (when $\epsilon \ne 0$), is a
Neimark-Saker bifurcation (also called secondary  Hopf), which
corresponds to two complex conjugate values of the fundamental matrix of
the ODE system crossing the unit  circle. The fundameltal matrix $H$ 
is defined as 
$dH/dt = J(x(t))H$ with $H(x, 0) = I$, with $J$ the Jacobian matrix of 
the ODE system and $I$ the identity.  The numerical software
CANDYS/QA \cite{FJ} has been used to track this bifurcation. For a given parameter set a bifurcation curve like in
Fig.~\ref{fig:stability} can be 
obtained as follows:  for a given driving velocity, one starts from a low
enough normal load $W_0$ in order to be in the steady sliding regime.
Once such a ``first point'' is indeed found by the software, one
varies $W_0$ only  (1-parameter continuation) until a bifurcation is
detected ($W_0=W_0^c$); one then follows this bifurcation curve by
further varying 
the driving velocity too (2-parameter continuation).

The procedure to detect the bifurcation has  been
automated. Starting from the parameter values determined in
\ref{sec:deterpar}, the critical values $W_0^c$ are determined and 
compared to the experimental ones ${W_0^c}_{\text {exp}}$. 
A systematic procedure
(Powell's method, as described in \cite{NUM}) is then used
that attemps to minimize 
$\sum_{V, \epsilon}[W_0^c(\rho\epsilon)-{W_0^c}_{\text {exp}}]^2$, 
with $\rho = \eff/\epsilon$, for a representative set of experimental data.  
The
set of parameters hence determined, which corresponds to a local minimum 
of the cost function, reducing it by a factor of fifteen~\footnote{When several critical values 
for $W_0$ can be detected, the
one retained in the evaluation of the cost function is the closest chosen among
the odd ones (first, third, etc.), corresponding to  a transition from steady sliding to
stick-slip when increasing $W_0$, while even ones correspond to a
transition from steading sliding to stick-slip.}, is given in
Table~\ref{tab:par}. 

The full  bifurcation curves are then determined as described above,
by a 2-parameter continuation. The results are shown in 
figures~\ref{fig:stability} and~\ref{fig:chichi} together with 
experimental data covering a range of $\eff$ values wider than the 
one involved in the adjustment procedure.

\section{Discussion}

\begin{figure}
\includegraphics[width=10cm]{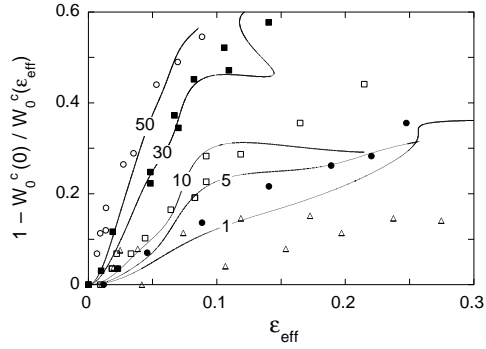}
\caption{
Reduced critical load {\it vs.} $\eff$ for different driving 
velocities: $V(\mu$m.s$^{-1})$ = $1\,(\vartriangle)$; $5\,(\bullet)$; 
$10\,(\square)$; $30\,(\blacksquare)$; 
$50\,(\circ)$.  The curves are the output of the numerical study (see 
section~\ref{sec:gigafit}), labeled with the corresponding velocities in 
$\mu$m.s$^{-1}$.
}
\label{fig:chichi}
\end{figure}

The curves displayed in Figs.~\ref{fig:stability} and \ref{fig:chichi} 
have been calculated with the optimized set of parameters. However, it 
is worth noting that the trial set yields numerical results in good qualitative 
agreement with the experimental data as well.
Namely, the main effect of normal load modulation, at least for 
moderate values of $\eff$, which is to stabilize sliding against 
stick-slip oscillations, is well reproduced by the SRF model. 
Moreover, the enhanced efficiency of the modulation on increasing the 
sliding velocity is correctly accounted for. 

The set of optimized data differs from the trial set essentially for 
three parameters, namely the elastic length $\lambda$, the saturation 
velocity $\Vsat$, and the ratio $\rho = \eff /\epsilon$ accounting for the air-cushion 
effect. 

The final value for $\lambda$ lies within the error 
bars estimated in \cite{PRE00}. 

As already mentioned in Sec.~\ref{sec:deterpar}, the large variation of  
$\Vsat$ during the optimization procedure is attributable to the 
fact that the crossover from a velocity-weakening regime to a strengthening 
one for steady sliding friction lies well above the upper experimental 
velocity, whence the goodness of the fit is only weakly sensitive to $\Vsat$.

The ratio $\rho = \eff /\epsilon$ was determined in \cite{PRE00} 
by comparison of the experimental shift of the steady friction level
$\Delta\mu_{0}$  at 120 Hz and the value predicted by the SRF model. The value 
taken in this reference was $\rho = 0.48$. 
Taking into account 
the error bars on $\Delta\mu_{0}$ one finds that the relative uncertainty 
on $\rho$ is about $\pm 20 \%$. The optimized value for this parameter lies 
therefore within this range. 

Thus, the SRF model with its set of parameters as determined from the 
dynamical study of the system under constant normal load is fully predictive 
as regards the sliding stability of the system under modulated load, 
at least for the values of $\eff$ probed by the data of 
Fig.~\ref{fig:stability}. In turn, the sensitivity of this experiment 
enabled 
us to refine the determination of the parameters. 

The quantitative overall agreement between the experimental data and the 
numerical curves in Fig.~\ref{fig:chichi} is excellent for, say, 
$\eff < 0.15$. Above this value the calculated curves tend to 
fold and correspond to a re-entrent stability 
diagram; namely, for given $\eff$ and $V$, increasing $K/W_{0}$ yields 
successive bifurcations from stick-slip to steady sliding then back to 
steady-sliding, etc. No experimental evidence of such an unexpected behaviour 
has been encountered so far. It is clearly the result of the non-linear 
coupling between the normal load modulation and the stick-slip 
oscillations. As such, it is expected to depend drastically on the 
{\it details} of the SRF laws. 
The importance of the terms which ultimately 
cut-off the logarithmic variations in the SRF laws has been stressed in 
several studies \cite{PRE95,JP1}. The existence of $\Vsat$, which accounts for a short time 
cut-off in the creep deformation of the load-bearing asperities 
\cite{PRB1}, yields one of these terms. It should be kept in 
mind that the SRF constitutive law~(\ref{eq:frico})  retains only the 
leading terms in the expansion of the friction force in powers of 
${\ln}(v)$. For instance,  Eq.~(\ref{eq:taborsrf}) 
with physically sounded expressions for $\sigma_{s}(v)$ and 
$\Sigma_{r}(\phi)$ would lead to terms of order 
$AB\ln(v)\ln(\phi)$ which, though negligible for most purposes, 
would probably affect the critical behaviour of the system under a 
strongly modulated load. 
For these reasons, we think that a full quantitative agreement between 
the experimental bifurcation at large 
\footnote {It has been shown in \cite{PRE00} 
that the relevant perturbation parameter is actually 
$\mu_{0}\eff /A \gg \eff$ which is already larger than 1 for $\eff = 
0.1$. 
}  
$\eff$  and the SRF model predictions would be illusive.

In addition, we have investigated numerically the effect of the extra term 
in Eq.~(\ref{eq:phidot})
proposed by Linker and Dieterich \cite{LD} and already discussed in 
\cite{PRE00}. No significant effect has been found at 120 
Hz, where we conclude that this term, if it exists as such, is not 
relevant to the present relatively high frequency study. 

\section{Conclusion}

The stability of a sliding system with a few degrees of freedom, 
submitted to a periodically modulated normal load, has been studied 
experimentally. The study clearly evidences the role of load 
modulation, even at moderate amplitude, as a stabilizer against stick-slip oscillations. 
Though this effect of vibrations is seemingly part of the empirical culture in 
mechanical engineering, it is the first time, as far as we know, that 
it is investigated experimentally. This effect should be of great interest in 
fault mechanics as well as in the control of precision structures. 
The results have been compared to the numerical predictions of a 
 model of the SRF type, relevant to multicontact friction at 
low velocities and low loads, including finite interfacial shear 
stiffness as a key parameter. Excellent quantitative agreement has 
been found as long as the amplitude of load modulation is restricted 
to about 10$\%$ of the dead load. 

Although, as discussed above, the main effect of the normal load 
modulation is stabilization,  the numerical study strongly suggests
that {\it destabilization} may also occur, due to the highly non-linear 
features of the model which also 
gives rise to rentrent stability diagram in Fig.~\ref{fig:stability}. 
More precisely, it can be seen in this figure that the $\eff = 0.18$ 
curve crosses the $\eff = 0.13$ one around $V = 7\,\mu$m.s$^{-1}$. 
For a $(V, K/W_{0})$ point slightly on the right of this crossing, 
in between the two curves, increasing $\eff$ would result in a bifurcation 
from stable sliding to stick-slip.  
This effect has not been observed directly so far, probably because 
it corresponds to small regions of the parameter space, strongly 
dependent on the value of the parameters.  
Clearly, this point would deserve further experimental study.  

\appendix
\section*{Linear stability analysis for $\epsilon = 0$}

The linear stability analysis of the SRF equations has been performed 
previously \cite{RR,PRE94}. However, difference 
between sliding velocity and the velocity of the center of mass of the 
slider was disregarded in these works. Since the interfacial stiffness is of paramount importance 
when the normal load is modulated at relatively high frequency, it is 
necessary to evaluate its role on the location of the bifurcation 
under constant load. Moreover, we will derive in this appendix expressions for the 
critical stiffness and the critical pulsation that hold for any state- 
and rate-dependent friction force. 
  
Let us consider a general expression: 
\begin{equation} 
F = W_{0}\mu(v, \phi)\ .
\end{equation}
The time-evolution of the age variable $\phi$ is ruled by: 
\begin{equation}
\label{eq:age}
\phi' = 1-\frac{v\phi}{D_{0}}\ .
\end{equation} 

When the slider is driven at constant velocity $V$, the 
steady sliding values of the dynamical variables are $v = V$ and 
$\phi = D_{0}/V$. 
We define: 
\begin{equation}
\label{eq:muvmuphi}
\left\lbrace\begin{array}{l}
\mu_{v} = \frac{\displaystyle{\partial\mu}}{\displaystyle{\partial\ln v}}(V, D_{0}/V)>0\\
\mu_{\phi} = \frac{\displaystyle{\partial\mu}}{\displaystyle{\partial 
\ln\phi}}(V, D_{0}/V)>0\ .\\
\end{array}\right.
\end{equation}
The position of the center of mass of the slider is $x(t)$ so that 
the elongation of the loading spring is $x-Vt$. At frequencies of 
interest the interfacial zone can be assumed massless and essentially 
elastic with a frequency independent, real stiffness $\kappa$ 
\cite{RSI}.  
The following relation thus holds: 
\begin{equation}
\label{eq:slipv}
v = x' + \frac{d}{dt}\left(\frac{M x'' - K(Vt-x)}{\kappa}\right)\ .
\end{equation}
We will make use in the following of the ratio $\eta$ of the loading 
spring 
stiffness $K$ to the equivalent stiffness of the loading spring in 
parallel with the interface $K\parallel\kappa$:
\begin{equation}
\eta = \frac{K}{K\parallel\kappa} = \frac{K+\kappa}{\kappa}\ .
\end{equation}
Finally, the dynamical equation for the motion of the slider reads:
\begin{equation}
\label{eq:rfd}
Mx'' = K(Vt-x)-W_{0}\mu(v, \phi)\ .
\end{equation}
The set of dynamical equations  
(\ref{eq:age},\ref{eq:slipv},\ref{eq:rfd}) is closed and can be 
linearized about the steady sliding state, setting:
\begin{equation}
\left\lbrace\begin{array}{ll}
x = Vt - F(V, D_{0}/V)/K +\delta x, & |\delta x|\ll F(V, D_{0}/V)/K \\
\phi = D_{0}/V +\delta \phi, & |\delta \phi|\ll D_{0}/V\ . \\
\end{array}
\right.
\end{equation}
The linearized system becomes:

\begin{widetext}
\begin{equation}
\label{eq:linsystem}
\left\lbrace \begin{array}{l}
M\delta x'' = 
-K\delta x 
- W_{0}\left[(\mu_{v}/V)\left(\eta \delta x'
 + \frac{M}{\kappa} \delta x'''\right)+(\mu_{\phi}V/D_{0})\delta\phi\right] \\
\delta\phi' = 
-\eta\delta x'/V - \frac{M}{K}  \delta x'''/V - \delta\phi 
V/D_{0}
\\
\end{array}
\right.
\end{equation}
\end{widetext}
The solutions are  the real parts of the complex 
$\tilde{\delta x} = \tilde{\delta x_{0}} \exp(i\Omega t)$ and 
$\tilde{\delta \phi} = \tilde{\delta \phi_{0}} \exp(i\Omega t)$
with $\Omega$ a complex number. 
Replacing into Eq. (\ref{eq:linsystem}) and writing the condition for 
non-trivial amplitudes $\tilde{\delta x_{0}}$ and 
$\tilde{\delta \phi_{0}}$, one finds the dispersion relationship
\begin{equation}
\label{eq:dispersion}
C_{4}\Omega^{4} + C_{3}\Omega^{3}+ C_{2}\Omega^{2}+ C_{1}\Omega + 
C_{0} = 0
\end{equation}
with:
\begin{equation}
\label{eq:coefs}
\left\lbrace
\begin{array}{l}
C_{0}= \frac{KV}{D_{0}}\\
C_{1}=i\frac{W_{0}}{D_{0}}\left[\frac{KD_{0}}{W_{0}}-\eta\left(\mu_{\phi}-\mu_{v}\right)\right]\\ 
C_{2} =-\left(\frac{MV}{D_{0}}+\frac{\eta\mu_{v} W_{0}}{V}\right) \\
C_{3} = -iM\frac{W_{0}}{\kappa D_{0}}\left[\frac{\kappa 
D_{0}}{W_{0}}+\left(\mu_{v}-\mu_{\phi}\right)\right]\\
C_{4} = M\mu_{v}\frac{W_{0}}{\kappa V}\ . \\
\end{array}
\right.
\end{equation}
The critical value of the control parameters and the critical 
pulsation $\Omega_{c}$ are obtained
by expressing that at the Hopf bifurcation (at least) 
one root of Eq. (\ref{eq:dispersion}) crosses the imaginary 
axis. Setting that $\Omega$ is purely imaginary and extracting the 
real and imaginary components from Eq. (\ref{eq:dispersion}) yield 
the requested values. 

Let us first solve for an infinitely stiff interface, {\it i.e.} for 
$\eta = 1$ and $\kappa\to\infty$.  It is then straightforward to find:
\begin{equation}
\label{eq:kcinert}
\frac{KD_{0}}{W_{0}^c} = \left(\mu_{\phi}-\mu_{v}\right)
\left(1+\frac{MV^{2}}{W_{0}^cD_{0}\mu_{v}}\right)
\end{equation}
and
\begin{equation}
\label{eq:pulsinert}
\frac{D_{0}\Omega_{c}}{V} = 
\sqrt{\frac{\mu_{\phi}-\mu_{v}}{\mu_{v}}}\ .
\end{equation}
These relations make sense only for $\mu_{\phi}-\mu_{v}>0$, {\it i.e.} when 
the steady sliding friction coefficient $\mu^{ss}(V)$ is velocity 
weakening: $\partial \mu^{ss}/\partial \ln V <0$. 
For the particular expression of $\mu(v, \phi)$ used in the numerical 
analysis, this reads:
\begin{equation}
\label{eq:weakslope}
-\frac{\partial \mu^{ss}}{\partial \ln V} = \mu_{\phi}-\mu_{v} = 
\frac{B}{1+V/V_{sat}}-A >0\ .
\end{equation}
Now, let us evaluate the contribution of the finite interfacial 
stiffness $\kappa$ to Eq. (\ref{eq:dispersion}) by estimating the 
order of magnitude of $c =|C_{4}\Omega_{c}^{4}/C_{0}|$ with $\Omega_{c}$ given by 
Eq. (\ref{eq:pulsinert}), {\it i.e.} $\Omega_{c}\simeq V/D_{0}$. 
This reads:
\begin{equation}
c \simeq \mu_{v}\frac {MV^{2}}{KD_{0}^{2}}
\end{equation}
where we have expressed that $W_{0}/(\kappa D_{0}) = \lambda /D_{0}\simeq 1$. 
One can estimate $c< 10^{-3}$ within the experimental velocity 
range, hence the fourth order term in Eq. (\ref{eq:dispersion}) can be 
safely discarded. 
Next, a finite $\kappa$ introduces perturbative terms in $C_{3}$ 
which are of order $\mu_{v},\,\mu_{\phi}\simeq 10^{-2}$, still well below 
the relative uncertainty on the experimental determination of the critical parameters. 
Since the correction to the drive stiffness due to the interfacial 
elastic
element is of order 
$\eta-1 = K/\kappa \simeq 10^{-2}$, it can be concluded that for 
the purpose of calculating the values of the critical parameters, the 
finite interfacial stiffness has no practical effect, and one can 
make use of Eqs. (\ref{eq:kcinert}) and (\ref{eq:pulsinert}). 

\begin{table}
\caption{
\label{tab:par}
Values of the SRF parameters. 
}
\begin{tabular}{|l|l|l|}
parameter & trial value & optimized value \\
\hline
$K$ (N.$\mu$m$^{-1})$ & 0.21 & \\
\hline
$M$ (kg) & 2.37 & \\
\hline
$\mu_{0}(V_{0}=1\,\mu$m.s$^{-1}$) & 0.33 &  \\
\hline
$A$ & $0.012 \pm 0.002$ & 0.0126 \\
\hline
$B$ & $0.023 \pm 0.002$  & 0.0241 \\
\hline
$D_{0}$ ($\mu$m) & $0.40 \pm 0.04$ & 0.402 \\
\hline
$\Vsat$ ($\mu$m.s$^{-1}$) & $280 \pm 70$ & 256 \\
\hline
$\lambda$ ($\mu$m) & $0.62 \pm 0.15$ & 0.56 \\
\hline
$\rho = \eff /\epsilon$ & $0.48 \pm 0.10$ & 0.43 \\
\hline
\end{tabular}
\end{table}

\begin{acknowledgments}
The authors would like to thank Christiane Caroli and Jim Rice for discussions 
and encouragement, as well as Wolfgang Jansen for prompt assistance 
with CANDYS/QA. 
\end{acknowledgments}

\newpage


\begin{thebibliography}{}

\bibitem{DIET} J. H. Dieterich, 1979, ``Modeling of Rock Friction 1.
Experimental Results and Constitutive Equations,'' {\it Journal of Geophysical
Research}, Vol. 84, pp. 2161--2168.

\bibitem{RR} J. R. Rice, and A. L. Ruina,
1983, ``Stability of Steady Frictional Slipping,''
{\it ASME Journal of Applied Mechanics}, Vol. 105, pp. 343--349.

\bibitem{RAB} E. Rabinowicz, 1965, ``Friction and Wear of Materials,''
Wiley, New York.

\bibitem{PRE94} F. Heslot, T. Baumberger, B. Perrin, B.
Caroli, and C. Caroli, 1994, ``Creep, Stick-Slip, and Dry Friction
Dynamics: Experiments and a Heuristic Model,'' {\it Physical Review E},
Vol. 49, pp.
4973--4988.

\bibitem{PRB2} T. Baumberger, P. Berthoud, and C. Caroli,
1999, ``Physical Analysis of the State- and Rate-Dependent Friction Law:
II. Dynamic Friction,'' {\it Physical Review B}, Vol. 60, pp.
3928--3939.

\bibitem{PRS01} O. Ronsin, and K. Labastie-Coueyrehourcq, 2001, ``State, Rate
and Temperature-Dependent Sliding
Friction of Elastomers,''
{\it Proceedings of the Royal Society (London) A},
Vol. 457, pp. 1277--1294.

\bibitem{PRB1} P. Berthoud, T. Baumberger, C. G'Sell, and J.-M. Hiver,
1999, ``Physical Analysis of the State- and Rate-Dependent Friction Law:
Static Friction,'' {\it Physical Review B}, Vol. 59, pp.
14313--14327.

\bibitem{DK} J. H. Dieterich, and D. Kilgore, 1994, 
``Direct observation of frictional contacts:  New insights for
state-dependent properties'', {\it Pure and Applied Geophysics}
Vol. 43, pp. 283--302. 

\bibitem{GW} J. A. Greenwood, and J. B. P. Williamson, 1966, ``Contact
of Nominally Flat Surfaces,''
{\it Proceedings of the Royal Society (London) A}, Vol. 295, pp. 300--319.

\bibitem{PRE95} 
T. Baumberger, C. Caroli, B. Perrin, and O. Ronsin, 1995,
``Nonlinear Analysis of the Stick-Slip Bifurcation in the
Creep-Controlled Regime of Dry Friction,'' {\it Physical Review E}, Vol.
51, pp. 4005--4010.

\bibitem{BT} F. P. Bowden and D. Tabor, 1950, ``Friction and 
lubrication of solids'', Clarendon, Oxford. 

\bibitem{DB} P. E. Dupont, and D. Bapna, 1994, ``Stability of Sliding
Frictional Surfaces with Varying Normal Force,'' {\it ASME Journal of
Vibration and Acoustics}, Vol. 116, pp. 237--242.

\bibitem{LD} M. Linker, and J. H. Dieterich, 1992, ``Effects of
Variable Normal Stress on Rock Friction: Observations and Constitutive
Equations,'' {\it Journal of Geophysical Research}, Vol. 124,
pp. 445--485.

\bibitem{PERF} H. Perfettini, J. Schmittbuhl, J. R. Rice, and M. Cocco,
2001, ``Frictional Response Induced by Time-Dependent Fluctuations of
the Normal Loading,'' {\it Journal of Geophysical Research}, in press.

\bibitem{RM} E. Richardson and C. Marone, 1999,
``Effects of Normal Stress Vibrations on Frictional Heating'',
{\it Journal of Geophysical Research},
Vol. 104,
pp. 28,859--28,878. 

\bibitem{PRE00} L. Bureau, T. Baumberger, and C. Caroli, 2000, ``Shear Response
of a Frictional Interface to a Normal Load Modulation,'' {\it  Physical
Review E}, Vol. 62, pp. 6810--6820.

\bibitem{RSI} T. Baumberger, L. Bureau, M. Busson, E. Falcon, and B.
Perrin, 1998, ``An Inertial Tribometer for Measuring Microslip Dissipation
at a Solid-Solid
Multicontact Interface,'' {\it Review of Scientific Instruments}, Vol.
69, pp. 2416--2420.

\bibitem{PRS} P. Berthoud, and T. Baumberger, 1998, ``Shear Stiffness of a
Solid-Solid Multicontact Interface,'' {\it Proceedings of the Royal Society
(London) A}, Vol. 454, pp. 1615--1634.

\bibitem{FJ} U. Feudel,  and W. Jansen, 1992, ``CANDYS/QA - a software
system for the qualitative analysis of nonlinear dynamical systems,''
{\it Int. J. Bifurcation \& Chaos}, Vol. 2,
pp. 773-794.  See also
http://www.agnld.uni-potsdam.de/~wolfgang/wolfgang.html

\bibitem{NUM} W. H. Press, S. A. Teukolsky, W. T. Vetterling, and
B. P. Flannery, 1992, {\it Numerical Recipes}, Cambridge University
Press. 

\bibitem{JP1} T. Baumberger and L. Gauthier, 1996, ``Relaxation at the Interface between 
Rough Solids under Shear,'' {\it J. Phys. I France}, Vol. 6, pp. 
1021--1025.

\end{thebibliography}
\end{document}